\documentstyle[12pt]{article}

\begin{document}

\newcommand{\nit}{\noindent}
\newcommand{\nl}{\newline}
\newcommand{\np}{\newpage}
\newcommand{\dsp}{\displaystyle}
\newcommand{\be}{\begin{equation}}
\newcommand{\ee}{\end{equation}}
\newcommand{\ba}{\begin{array}}
\newcommand{\ea}{\end{array}}

\begin{titlepage}
\title{{\bf Relativistic Dynamics of Spin in Strong External Fields \footnote{
            Lecture presented at the 4th Hellenic School on Elementary Particle
            Physics, Corfu, Sept.\ 1992} }}

{\bf
\author{ J.W.\ van Holten \\ NIKHEF-H \\ P.O.\ Box 41882 \\ 1009 DB
         Amsterdam NL}
}

\date{}

\maketitle

\begin{abstract}
\noindent
The dynamics of relativistic spinning particles in strong external
electromagnetic or gravitational fields is discussed. Spin-orbit coupling
is shown to affect such relativistic phenomena as time-dilation and
perihelion shift. Possible applications include muon decay in a magnetic field
and the dynamics of neutron stars in binary systems.
\end{abstract}

\thispagestyle{empty}
\end{titlepage}

\pagestyle{plain}

\noindent
{\bf 1.\ Introduction} \newline

\noindent
The classical mechanics of spinning non-relativistic particles in external
fields predicts several interesting effects as a result of spin-orbit and/or
spin-spin coupling, which cause their behaviour to deviate from that of scalar
point particles under similar conditions \cite{T}-\cite{K}. One such effect is
that relativistic time dilation can have a {\em dynamical} component
\cite{WP,JH1}, in addition to the universal kinematic time dilation which
disappears in any inertial frame with respect to which the particle is at rest.
Such a non-universal dynamical effect can arise both in special and in general
relativity \cite{JH2,JH3}.

It is the purpose of this review to present some simple examples of these
dynamical phenomena, derived for spinning particles which satisfy equations of
motion of the type proposed by Frenkel \cite{F} and Bargmann, Michel and
Telegdi \cite{BMT} for motion in an electro-magnetic field, or a generalisation
of these equations to motion in a gravitational field \cite{P,JH1,JH3,K}.
However, in order to keep the analysis simple I have generally neglected
corrections to the motion of spins compensating for a non-canonical value of
the
gyromagnetic factor, assuming where necessary that $g = 2$ as for an electron
or other Dirac particle. Also, I will restrict the analysis below to effects
which depend only linearly on the spin, arising essentially from spin-orbit
coupling, whilst non-linear spin effects such as associated with spin-spin
coupling are disregarded. In view of the smallness of spin-effects this
approximation seems well-justified and it is not to be expected that these
restrictions alter the conclusions in a significant way, at least not
qualitatively. After all, in atomic physics hyperfine splitting is generally
small compared to fine-splitting, which in turn tends to be small compared to
the transition energies between states of different principal quantum numbers.
And at the other extreme, the ratio of intrinsic to total angular momentum
of a rapidly spinning neutron star orbiting a heavy companion is typically of
the order of $10^{-3}$ or smaller. Finally, since the effects predicted are not
special to classical particles, but have quantum mechanical counterparts, one
may feel quite confident that the considerations presented below are of fairly
general physical interest.
\nl\nl

\noindent
{\bf 2.\ Charged point particle in a magnetic field} \newline

\noindent
To show how relativistic effects of spin in an external field may arise, I
consider first the motion of a spinless charged point particle in a constant
magnetic field, neglecting the back reaction of the motion of the charge on the
field. The Lorentz force law provides the classical equation of motion for the
particle:

\be
\dsp{ \frac{d}{dt}\, \frac{m \vec{v}}{\sqrt{1 - \vec{v}^{\, 2}/c^{2}}}\, =\,
      \frac{q}{c}\, \vec{v} \times \vec{B}. }
\label{1}
\ee

\nit
Taking the direction of the field as the $z$-axis, the motion can be decomposed
into a linear motion with constant velocity $v_{z}$ parallel to the field, and
a circular motion with frequency

\be
\omega\, =\, \frac{qB}{mc}\, \sqrt{1 - \vec{v}^{\, 2}/c^{2}},
\label{2}
\ee

\nit
(the cyclotron frequency) perpendicular to the field. Thus the solution of
eq.(\ref{1}) is

\be
\vec{v}\, =\, \left( \omega r \sin \omega t, \omega r \cos \omega t, v_{z}
              \right).
\label{3}
\ee

\nit
Elimination of $\omega$ then gives

\be
\ba{lll}
\dsp{ 1\, -\, \frac{\vec{v}^{\, 2}}{c^{2}} }& = & \dsp{
      1\, -\, \frac{v_{z}^{2}}{c^{2}}\, -\, \frac{\omega^{2} r^{2}}{c^{2}} }\\
  &  &  \\
  & = & \dsp{ \frac{ 1 - v_{z}^{2}/c^{2} }{ 1 + (qBr/mc^{2})^{2} }. }\\
\ea
\label{4}
\ee

\nit
To interpret this formula we associate an effective magnetic moment with the
circulating charge, as induced by the external field:

\be
\vec{\mu}\, =\, \frac{q}{2mc}\, \vec{r} \times \frac{m\vec{v}}{
                \sqrt{1 - \vec{v}^{\, 2}/c^{2}} }.
\label{5}
\ee

\nit
In agreement with Lenz's law, it is directed opposite to the field and its
component in this direction has magnitude

\be
\frac{\vec{\mu} \cdot \vec{B}}{mc^{2}}\, =\, - \frac{1}{2}\, \left(
   \frac{qBr}{mc^{2}} \right)^{2}.
\label{6}
\ee

\nit
Combining eqs.(\ref{4}) and (\ref{6}) one finally obtains

\be
\frac{1}{1 - \vec{v}^{\, 2}/c^{2}}\, =\,
        \frac{1 - 2 \vec{\mu} \cdot \vec{B}/mc^{2}}{1 - v_{z}^{2}/c^{2}}.
\label{7}
\ee

\nit
As a result the relativistic expression for the energy can be written as

\be
E\, =\, mc^{2}\, \sqrt{ \frac{1 - 2 \vec{\mu} \cdot \vec{B}/mc^{2}}{
   1 - v_{z}^{2}/c^{2}} },
\label{8}
\ee

\nit
or equivalently

\be
E^{2}\, =\, m^{2} c^{4}\,+\, p_{z}^{2} c^{2}\, -\, 2 \vec{\mu} \cdot \vec{B}\,
            m c^{2}.
\label{9}
\ee

\nit
If the particle is a pion, for example, its mean life time is changed by the
relativistic time dilation factor

\be
\Delta t\, =\, \Delta \tau\, \sqrt{ \frac{1 - 2 \vec{\mu} \cdot
\vec{B}/mc^{2}}{
               1 - v_{z}^{2}/c^{2}} }.
\label{10}
\ee

\nit
Here $\tau$ denotes the proper time, and $t$ the laboratory time measured
in the observers rest frame. To an observer at a distance $R \gg r$, for whom
the circular motion is not resolved, it seems that the pion moves linearly
parallel to the field with a magnetic moment $\mu_{z} = - q^{2} r^{2} B/2
mc^{2}$. For this observer the mean life time depends both on the velocity
(along the field direction) {\em and} on the energy $\vec{\mu} \cdot \vec{B}$
associated with the magnetic moment. In particular, the time dilation does not
vanish in the limit in which the translational velocity $v_{z}$ goes to zero.
In
the following similar results will be derived for particles with an intrinsic
magnetic moment.
\nl\nl

\nit
{\bf 3.\ Classical dipoles} \nl

\nit
Consider a charged particle with an intrinsic electric dipole moment $\vec{d}$
and a magnetic dipole moment $\vec{\mu}$. Like the electric and magnetic fields
themselves they can be assembled into a covariant anti-symmetric tensor, which
we associate with the spin (for $g = 2$), as follows:

\be
\frac{q}{mc}\, S^{\mu\nu} =\, \left( \ba{cccc}
                              0 & -c d_{x} & -c d_{y} & -c d_{z} \\
                              c d_{x}  & 0 & \mu_{z}  & -\mu_{y} \\
                              c d_{y}  & -\mu_{z} & 0 &  \mu_{x} \\
                              c d_{z}  &  \mu_{z} & -\mu_{x} & 0 \\
                                      \ea \right).
\label{11}
\ee

\nit
Equivalently, we may associate the electric and magnetic dipole moments with
the Lorentz four-vectors

\be
\ba{lll}
D_{\mu} & = & \dsp{ \frac{q}{mc^{3}}\, S_{\mu\nu}\, \dot{x}^{\nu}, }\\
  &  &  \\
M_{\mu} & = & \dsp{ \frac{q}{2mc^{2}}\, \varepsilon_{\mu\nu\kappa\lambda}\,
            \dot{x}^{\nu}\, S^{\kappa\lambda}, }\\
\ea
\label{12}
\ee

\nit
where the overdot denotes a proper-time derivative. Clearly, in the rest frame
these four-vectors reduce to the electric and magnetic three-vectors $(\vec{d},
\vec{\mu})$. However, in a general inertial frame the electric and magnetic
dipole moments mix, and a particle with vanishing electric dipole moment in the
rest
frame nevertheless develops one with respect to a moving observer.

The energy of the dipoles in an external field can now be written as

\be
\vec{d} \cdot  \vec{E}\, +\, \vec{\mu} \cdot \vec{B}\, =\, \frac{q}{2mc}\,
     S^{\mu\nu} F_{\mu\nu}.
\label{13}
\ee

\nit
{}From the right-hand side of this equation we infer, that this energy is a
relativistic invariant.

We now propose the following equations of motion for a classical particle with
magnetic and/or electric dipole moments:

\be
\ba{lll}
m \ddot{x}^{\mu} & = & \dsp{ q F^{\mu}_{\:\:\nu}\, \dot{x}^{\nu}\, +\,
  \frac{q}{2mc}\, S^{\kappa\lambda}\, \partial^{\mu}\, F_{\kappa\lambda}, }\\
  &  &  \\
\dot{S}^{\mu\nu} & = & \dsp{ \frac{q}{m}\, \left[ F, S \right]^{\mu\nu}. }\\
\ea
\label{14}
\ee

\nit
The right-hand side of the last equation is the ordinary commutator of the two
tensors $F$ and $S$ considered as $4 \times 4$ matrices. These equations have
the following physical interpretation: the first term on the right-hand side of
the first equation represents the usual Lorentz force; the second term is a
direct covariant extension of the Stern-Gerlach force, which results from the
coupling of the dipole moment to the gradient of the field. The last equation
is
the expression of Frenkel \cite{F} and Bargmann, Michel and Telegdi \cite{BMT}
for the precession of a dipole moment in an external field in the limit $g =
2$.

The first integral of motion obtained from these equations leads to the
following relativistic energy-momentum relation:

\be
(p_{\mu} - q A_{\mu})^{2}\, -\, \frac{q}{c}\, S^{\mu\nu} F_{\mu\nu}\, +\,
m^{2} c^{2}\, =\, 0.
\label{15}
\ee

\nit
This replaces the usual mass-shell condition and reduces in the
non-relativistic
limit to

\be
E\, =\, mc^{2}\, +\, \frac{1}{2m}\, (\vec{p} - q \vec{A})^{2}\, +\, q \phi\,
 - ( \vec{d} \cdot \vec{E} + \vec{\mu} \cdot \vec {B} )\, +\, ...
\label{16}
\ee

\nit
Here $q \phi$ is the electrostatic energy of the charge. The full solution of
the energy-momentum relation and the equations of motion is \cite{JH1,JH2}

\be
\ba{lll}
\vec{p}\, -\, q \vec{A} & = & \dsp{ m \vec{v}\, \sqrt{ \frac{1 - 2(\vec{d}
\cdot
    \vec{E} + \vec{\mu} \cdot \vec{B})/mc^{2}}{1 - \vec{v}^{\, 2}/c^{2}} }, }\\
 & & \\
E\, -\, q \phi & = & \dsp{ mc^{2}\, \sqrt{ \frac{1 - 2(\vec{d} \cdot
    \vec{E} + \vec{\mu} \cdot \vec{B})/mc^{2}}{1 - \vec{v}^{\, 2}/c^{2}} }, }\\
\ea
\label{17}
\ee

\nit
For the relativistic time dilation we now obtain the result

\be
\Delta t\, =\, \left( \frac{E - q \phi}{mc^{2}}\right)\, \Delta \tau\, =\,
               \Delta \tau\, \sqrt{ \frac{1 - 2(\vec{d} \cdot
    \vec{E} + \vec{\mu} \cdot \vec{B})/mc^{2}}{1 - \vec{v}^{\, 2}/c^{2}} }.
\label{18}
\ee

\nit
This agrees completely with our previous result for the motion of a point
particle in a magnetic field, if we replace the induced magnetic moment by the
intrinsic dipole moments. Note that again a time dilation factor remains
associated with the energy of the dipole in an external field even in the
limit of zero translational velocity.
\nl\nl

\nit
{\bf 4.\ Comparison with the Dirac equation} \nl

\nit
Now compare our classical equations with the quantum theory of spinning
particles, as represented by the Dirac equation for a particle in an external
field:

\be
\left( \gamma \cdot {\cal D}\, +\, \frac{mc}{\hbar} \right)\, \Psi\, =\, 0,
\label{19}
\ee

\nit
where the external field is taken to be electro-magnetic and the interaction
with the Dirac particle is described by minimal coupling:

\be
{\cal D}_{\mu}\, =\, \partial_{\mu}\, -\, \frac{i}{\hbar}\, q A_{\mu}.
\label{20}
\ee

\nit
Multiplying the Dirac equation by the operator $(- \gamma \cdot {\cal D} +
mc/\hbar)$ and working out the algebra of Dirac matrices and covariant
derivatives, we get a generalised Klein-Gordon equation

\be
\left( - {\cal D}_{\mu}^{2}\, +\, \frac{i}{\hbar}\, q \sigma^{\mu\nu}
 F_{\mu\nu}\, +\, \frac{m^{2}c^{2}}{\hbar^{2}} \right)\, \Psi\, =\, 0.
\label{21}
\ee

\nit
This corresponds exactly to the classical equation (\ref{15}) if make the
operator correspondence

\be
p_{\mu} \rightarrow - i \hbar \partial_{\mu}, \hspace{3em}
S^{\mu\nu} \rightarrow - i \hbar \sigma^{\mu\nu}.
\label{22}
\ee

\nit
Therefore our classical equations of motion (\ref{14}) may indeed be taken
to represent the classical limit of the Dirac equation, provided we take the
particle to have vanishing electric dipole moment in its rest frame far away
from external fields.

At this point it is however necessary to issue a warning: although it is
perfectly permissable to impose the vanishing of the intrinsic electric dipole
moment as an initial condition, its preservation in time is violated in
non-homogeneous fields by terms quadratic in the spin, unless special
conditions
on the spin variables are imposed \cite{JH1}. However, as stated in the
introduction, all such effects quadratic in the spin are neglected here because
of their numerical smallness. \nl\nl

\nit
{\bf 5.\ Muon decay} \nl

\nit
As an application of our results so far we consider $\beta$-decay of a muon in
an external field:

\be
\mu\, \rightarrow\, e \bar{\nu}_{e} \nu_{\mu}.
\label{23}
\ee

\nit
The mean life time of the free muon $\tau_{\mu}$ is accurately known to be
$2.19703 \times 10^{-6}$ sec. Given a muon mass of $105.7$ MeV, the change in
the life time we predict by placing the particle in an external magnetic field
of magnitude $B$ is

\be
\frac{\delta \tau}{\tau_{\mu}}\, =\, - \frac{\vec{\mu} \cdot \vec{B}}{mc^{2}}\,
   =\, 0.28 \times 10^{-14} \times B,
\label{24}
\ee

\nit
with $B$ measured in Tesla. In order to obtain a significant variation in the
mean life time we therefore need fields of at least $5 \times 10^{9}$ T. This
corresponds to a magnetic energy of

\be
\vec{\mu} \cdot \vec{B}\, \geq 1\, \mbox{keV}.
\label{25}
\ee

\nit
Such magnetic interaction energies are not unrealistic; for example, the
hyperfine splitting of muon energy levels in the field of high-$Z$ atomic
nuclei like Nb or Bi are of the order of 3 -- 5 keV. Practical difficulties in
measuring the corresponding change in life time are the result of competing
effects like muon-capture by the nucleus, and of the short life-time of the
higher state in the hyperfine doublet: the rate of transition to the
groundstate, an electromagnetic process, is much higher than the rate of muon
decay, which is a weak interaction process. A reliable measurement would
require
long-lived high population density in the excited state of the doublet, so as
to
allow accurate comparison of the relative life times between the two states.

Another place where very strong magnetic fields are encountered is near neutron
stars. It would be very interesting if it were possible to extract information
about unstable particle decay from such distant sources. \nl\nl

\nit
{\bf 6.\ Gravitational interactions} \nl

\nit
Next I turn to the case of a spinning particle moving in a gravitational
background field. Again we can make an inspired guess for the classical
equations of motion based upon correspondence with the Dirac equation. The
simplest conjecture \cite{JH1,JH3}, which disregards terms quadratic in the
spin
\cite{K}, is

\be
\ba{rcl}
\dsp{ m \frac{D^{2} x^{\mu}}{D\tau^{2}} } & = & \dsp{
   \frac{1}{2}\, S^{\kappa\lambda}R_{\kappa\lambda\:\:\nu}^{\:\:\:\:\:\mu}
   \dot{x}^{\nu}, }\\
  &  &  \\
\dsp{ \frac{D S^{\mu\nu}}{D\tau} } & = & 0. \\
\ea
\label{26}
\ee

\nit
Here $R_{\mu\nu\kappa}^{\:\:\:\:\:\:\:\:\lambda}$ is the Riemann curvature
tensor. Note that the form of the first equation is formally very similar to
the
Lorentz force law, with the electromagnetic field strength replaced by the
space-time curvature and the scalar charge $q$ replaced by the tensorial
coupling parameter $S^{\mu\nu}$, which is taken here to be covariantly
constant.
No additional Stern-Gerlach term has been introduced since it would be
quadratic in the spin.

As a first consequence one obtains as a first integral of motion the
mass-shell condition

\be
m^{2} g_{\mu\nu}\, \dot{x}^{\mu} \dot{x}^{\nu}\, +\, m^{2} c^{2}\, =\, 0.
\label{27}
\ee

\nit
As shown below, dynamical spin-dependent effects are now caused by spin
dependence of the solution $\vec{x}(t)$ for the orbit of the particle.

I will confine the discussion here to the simplest relevant physical situation
where these equations might apply: the motion of a spinning particle in the
field of a large spherically symmetric mass, like a star or a black hole
\cite{JH3}. In this case the space-time geometry is described by the
Schwarzschild solution of the Einstein equations:

\be
ds^{2}\, =\, - \left( 1 - \frac{\alpha}{r} \right) dt^{2}\, +\,
               \frac{1}{\left( 1 - \frac{\alpha}{r} \right)}\, dr^{2}\,
               +\, r^{2} d\theta^{2}\, +\, r^{2} \sin^{2} \theta\, d\phi^{2},
\label{28}
\ee

\nit
with $\alpha = 2 GM$, $M$ being the total central mass and $G$ Newton's
gravitational constant.

The solution of the equation of motion in the absence of spin is well-known:
orbital angular momentum is conserved, the motion is planar and, for bound
states, the orbit is to high accuracy an ellipse, with a precession of the
perihelion of magnitude

\be
\Delta \phi\, =\, \frac{6 \pi GM}{k}\, \left( 1 + (18 + e^{2}) \frac{GM}{4k}
                                              + ... \right).
\label{29}
\ee

\nit
In this equation $e$ is the eccentricity of then ellipse, and $k$ the
semi-latus
rectum \cite{Wb}.

The corrections to first order in the spin now amount to the following: instead
of conservation of orbital angular momentum $\vec{L}$, we now only have
conservation of {\em total} angular momentum $\vec{J} = \vec{L} + \vec{S}$. If
$\vec{L}$ and $\vec{S}$ are not parallel, they will both precess around
$\vec{J}$ in such a way as to keep $\vec{J}$ constant. Hence in general the
motion is no longer planar. Of course, if $|S| \ll |L|$ the motion is still
approximately planar. If $\vec{S}$ and $\vec{L}$ are parallel, this is an exact
result. The orbit will again be an approximate ellipse, but the rate of
precession of the perihelion is changed by spin-orbit coupling. One finds
\cite{JH3}

\be
\Delta \phi\, =\, \frac{6 \pi GM}{k}\, \left( 1 + \Delta + (18 + e^{2})
                  \frac{GM}{4k} + ... \right).
\label{30}
\ee

\nit
where, taking $\vec{L}$ in the $z$-direction, $\Delta = S_{z}/L$.

Another constant of motion in a static space-time geometry, like the
Schwarzschild metric, is the space-time energy (the time component of the
canonical energy-momentum 4-vector). For a spinning particle it is of the
form

\be
E\, =\, mc^{2} \left( 1 - \frac{\alpha}{r} \right)\, \frac{dt}{d\tau}\,
        -\, \frac{\alpha}{2r^{2}}\, S^{rt}.
\label{31}
\ee

\nit
Now for motion in a plane we have separate conservation of orbital and spin
angular momentum; hence

\be
L_{z}\, =\, m r^{2} \dot{\phi}
\label{32}
\ee

\nit
is a constant of motion. In addition, absence of the electric dipole components
of the spin gives

\be
S^{\mu\nu} \dot{x}_{\nu}\, =\, 0 \rightarrow S^{rt}\, =\, \frac{\vec{L} \cdot
\vec{S}}{Er}.
\label{33}
\ee

\nit
Combining these results we arrive at the following formula for time-dilation
in a Schwarzschild geometry

\be
dt\, =\, \frac{d\tau}{\left(1 - \frac{\alpha}{r}\right)}\, \frac{E}{mc^{2}}\,
         \left( 1 + \frac{\alpha \vec{L} \cdot \vec{S} }{2 E^{2} r^{3}}
\right).
\label{34}
\ee

\nit
Thus there is the usual universal gravitational redshift, given by the
first term inside the parentheses, and in addition a dynamical non-universal
redshift proportional to the spin-orbit coupling.

These examples show that the phenomena we encountered for the motion of
spinning
particles in an electro-magentic field generalize to the case of gravitational
fields. \nl\nl

\nit
{\bf 7.\ The binary pulsar}\nl

\nit
I can not predict precisely to what extend our results are valid for general
spinning particles, which do not have the canonical value of the gyromagnetic
factor. However, assuming that the above results are valid at least
qualitatively in the general case, we may estimate the order of magnitude of
the
various effects for a system like the binary pulsar PSR 1913 + 16. This is
a binary system including a rapidly spinning neutron star, the radio signal of
which has been studied for many years, giving very precise information about
its orbital parameters. The rapid spin of the neutron stars make them the only
known macroscopic objects whose spin has to be treated relativistically.

For a spherical mass $m$ with radius $R$ the moment of inertia is $I = 2/5\,
mR^{2}$, and the intrinsic angular momentum

\be
S\, =\, I \omega\, =\, \frac{2}{5}\, m R^{2} \omega.
\label{35}
\ee

\nit
With an orbital angular momentum given by eq.(\ref{32}) the ratio of the
two quantities becomes

\be
\Delta\, =\, \frac{S}{L}\,
         =\, \frac{2}{5}\, \frac{R^{2}\omega}{r^{2} \dot{\phi}}.
\label{36}
\ee

\nit
I have used here a non-relativistic expression for $S$, but this should be
allowed on order to make an order-of-magnitude estimate as it provides an upper
limit.

Now for PSR 1913 + 16 the respective quantities are approximately

\be
\ba{ll}
R\, =\, 2.5\, \times 10^{4}\, \mbox{m}, &
                              \omega\, =\, 106.4\, \mbox{rad sec$^{-1}$}, \\
  & \\
r\, =\, 6.2 \times 10^{8}\, \mbox{m}, &
               \dot{\phi}\, =\, 2.252 \times 10^{-4}\, \mbox{rad sec$^{-1}$}.\\
\ea
\label{37}
\ee

\nit
As a result we find $\Delta = S/L \approx 0.35 \times 10^{-3}$.

Finally we substitute this estimate into the equations for the perihelion shift
and the redshift formula. For the binary pulsar, the next-to-leading order
result for the perihelion shift of a spinless mass point is determined
by the quantity

\be
18 \frac{GM}{4k}\, \approx 0.23 \times 10^{-4},
\label{38}
\ee

\nit
cf.\ eq.(\ref{29}). This is to be compared with $\Delta$ in computing the
precession of the perihelion as in eq.(\ref{30}). Clearly $\Delta$ is an order
of magnitude {\em larger} than the next-to-leading order general relativistic
correction. Given the accuracy of the measurement of the orbit of the binary
pulsar, this effect seems within the limits of observation.

For the redshift effect, we have to compute the spin-orbit term

\be
\frac{\alpha \vec{L} \cdot \vec{S}}{2 E^{2} r^{3}}\, \approx\,
 1.4 \times 10^{-12}\, \Delta.
\label{39}
\ee

\nit
Clearly, for the value of $\Delta$ in the order of $10^{-3}$ as quoted above
this is completely negligeable compared to unity. Thus the spin-dependent
non-universal redshift seems unobservable, at least for this system.
\nl\nl

\nit
{\Large {\bf Acknowledgement}}
\vspace{.5cm}

\nit
It is a pleasure to thank Rachel Rietdijk for many discussions and her
collaboration in studying the motion of spinning particles in a gravitational
field.
\vfill
\np

\vfill

\vspace{.5cm}

\begin{thebibliography}{99}

\bibitem{T} L.H.\ Thomas, Nature 117 (1926), 514
\bibitem{F} J.\ Frenkel, Z.\ Physik 37 (1926), 243
\bibitem{WP} W.\ Petry, Rev.\ Roum.\ de Phys.\ 21 (1976), 289
\bibitem{JH1} J.W.\ van Holten, Nucl.\ Phys.\ B356 (1991), 3
\bibitem{JH2} J.W.\ van Holten, Physica A182 (1992), 279
\bibitem{JH3} R.H.\ Rietdijk and J.W.\ van Holten, preprint NIKHEF-H/92-09;
              Class.\ Quantum Grav.\ (to appear)
\bibitem{BMT} V.\ Bargmann, L.\ Michel and V.L.\ Telegdi, Phys.\ Rev.\ Lett.\
              2 (1959), 435
\bibitem{P} A.\ Papapetrou, Proc.\ Roy.\ Soc.\ A209 (1951), 248
\bibitem{K} I.B.\ Khriplovich, JETP 69 (1989), 217
\bibitem{Wb} S.\ Weinberg, {\em Gravitation and Cosmology\/}\, (J.\ Wiley,
1972)

\end{thebibliography}
\end{document}